\documentstyle[12pt]{article}
\newcommand{\be}{\begin{equation}}
\newcommand{\ee}{\end{equation}}
\newcommand{\bea}{\begin{eqnarray}}
\newcommand{\beas}{\begin{eqnarray*}}
\newcommand{\eea}{\end{eqnarray}}
\newcommand{\eeas}{\end{eqnarray*}}
\newcommand{\ba}{\begin{array}}
\newcommand{\ea}{\end{array}}
\renewcommand{\thefootnote}{\fnsymbol{footnote}}

\begin{document} 
\begin{titlepage} 
\begin{center}

{\Large\bf Neutrino Mixing and the Minimal 3-3-1 Model}

\medskip

\normalsize 
{\bf  A. Gusso$^{a}$, C. A. de S. Pires$^b$, and  P. S. Rodrigues da Silva$^a$}
\vskip .2cm
$^a$Instituto de
F\'{\i}sica Te\'{o}rica, Universidade Estadual Paulista,\\
 Rua Pamplona 145, 01405-900 S\~{a}o Paulo - SP, Brazil.
\vskip .1cm
$^b$ Departamento de F\'{\i}sica, Universidade Federal da
Para\'{\i}ba,\\
 Caixa Postal 5008, 58051-970, Jo\~ao Pessoa - PB
\vskip .2cm
\end{center}

\begin{abstract}
In the minimal 3-3-1 model charged leptons come in a non-diagonal
basis. Moreover the Yukawa interactions of the model
lead to a non-hermitian  charged lepton mass matrix. In other words,  the minimal 3-3-1 model 
presents a very complex lepton mixing. In view of this we check rigorously if the
possible textures of the lepton mass matrices allowed by the minimal 3-3-1 model can lead or not 
to the neutrino  mixing  required by the recent experiments in neutrino oscillation.

\end{abstract} 
\renewcommand{\thefootnote}{\arabic{footnote}} \end{titlepage} 

\section{Introduction}

An alternative to the Standard Model (SM) of electro-weak
interactions, the so called $3-3-1$ model, was proposed some
years ago~\cite{viceframp}, and one of its interesting features
is that neutrino are massive from the outset. However, at the
time the model was proposed there was no indication that
neutrinos were massive, and in order to avoid such a mass term
some fine tuning was necessary\cite{footpisano}.

In the last few years we witnessed a dramatic change in the
scenario of particle physics and the most recent experimental results, including
SNO~\cite{SNO}, K2K~\cite{K2K}, and KamLAND~\cite{KamLAND} data, are corroborating the
hypothesis of neutrino oscillation~\cite{BahcallConcha}, which
implies the existence of at least two massive neutrinos.  In view
of this the chief question now refers to the smallness of the
neutrino masses, and in the context of $3-3-1$ model, can these
masses be fairly accommodated? It was shown in Ref.~\cite{joshi}
that in fact they can. The model dispose of a type II seesaw
mechanism which provides small masses to neutrinos. This could
put an end point on the subject if the size of neutrinos masses
were the only issue. Nevertheless, the experiments are showing not
only that neutrinos have a tiny mass, but are also strongly
suggesting the pattern in which they mix~\cite{SNO,K2K}. The
proposal of this work is to check if the minimal 3-3-1 model can
explain the pattern of neutrino mixing in accordance with the
experiments.

There are two reasonable arguments to pursue this check. The
first one concerns its absence in literature, and it is
justifiable to look for consistency with experiment. The second
one is related to the interplay that exists between neutrino and
charged lepton mixing, that arises naturally inside  the minimal
3-3-1 model. Such interplay exists because neutrinos and charged lepton
masses both have the same origin: a sextet of scalars. Moreover
the charged lepton mass also receive contributions of a triplet
of scalars\cite{footpisano}. This implies that the charged leptons cannot be simply
assumed to lie on a diagonal basis. Consequently, in the minimal
3-3-1 model the charged leptons are non-trivially interconnected
to the neutrino physics and play some role in it, and we should
be able to know better how this connection is realized under the
light of the recent results on neutrino oscillation.

We organize this work as follows. First, in Sec.~\ref{sec1}, we
introduce the necessary ingredients of the model, presenting the
texture of neutrino and charged lepton mass matrix of the model.
 Then, in Sec.~\ref{sec2}, we examine the distinct
possibilities for the lepton mixing matrices suggested by the results of
neutrino oscillation. Finally, in Sec.~\ref{sec3}, we present our
conclusions.

\section{Lepton sector of the minimal 3-3-1 model}
\label{sec1}

In this section, we are going to present the ingredients
necessary to obtain the mass matrix for neutrinos and charged
leptons in the minimal $3-3-1$ model. The gauge group under
consideration is $SU_c(3)\otimes SU_L(3)\otimes U_X(1)$, which
spontaneously breaks to the usual $SU_c(3)\otimes SU_L(2)\otimes
U_Y(1)$. In its minimal version, the scalar
content that interacts with the lepton sector is composed by a
triplet, $ \eta =( \eta^0 , \eta^-_1 , \eta^+_2)^T \sim({\bf
1},{\bf 3}, 0)$  and a sextet $S$~\cite{footpisano},
\bea 
S=\left(
\begin{array}{ccc}
 \sigma^0_1 & h_1^{-} & s_2^{+} \\
 h_1^{-} & H_1^{--} & \sigma^0_2 \\
 h_2^{+} & \sigma^0_2 & H_2^{++}
\end{array}
\right)\sim({\bf 1},{\bf 6},0)\,,
 \label{sextet}
 \eea
where we have included their assigned transformation properties
under the gauge group $SU_c(3)\otimes SU_L(3)\otimes U_X(1)$.

The leptons are arranged in the fundamental representation of the
gauge group $SU_L(3)$, and for each family we have,
\bea \Psi_{a_L} =\left(
\begin{array}{c}
  \nu_a \\
  e_a \\
  e_a^c
\end{array}\right)_L\sim({\bf 1},{\bf 3},0)\,,
\label{leptons} \eea
where $a=1,\,2,\,3$ labels the different families.

The Yukawa interactions that generate the lepton masses are given
by,
\be {\cal L}^Y_l=\frac{1}{2}G_{ab}\overline{ (\Psi_{aL})^c}
S^*\Psi_{b_L}+ \frac{1}{2}F_{ab}\epsilon^{i j k}
\overline{(\Psi_{iaL})^c} \Psi_{jbL} \eta^*_k + \mbox{H.c.}
\label{yukawa} \ee
When the neutral scalars develop their respective vacuum
expectation values, $<\sigma^0_1>= v_{\sigma_1}$,
$<\sigma^0_2>=v_{\sigma_2}$  and $<\eta^0>= v_{\eta}$, neutrinos
and charged leptons acquire the following mass matrices,
\bea &&M_\nu= \left (
\begin{array}{lcr}
G_{11} & G_{12} & G_{13} \\
G_{21} & G_{22} & G_{23} \\
G_{31} & G_{32} & G_{33}
\end{array}
\right ) v_{\sigma_1}\, \label{neutmass1} \eea
and \bea &&M_l= \left (
\begin{array}{lcr}
\,\,\,\,\,\,\,\,\,G_{11}v_{\sigma_2} &
(G_{12}v_{\sigma_2}+F_{12}v_\eta )&
(G_{13}v_{\sigma_2}+F_{13}v_\eta ) \\
(G_{21}v_{\sigma_2}-F_{12}v_\eta ) & G_{22}v_{\sigma_2} &
(G_{23}v_{\sigma_2}+F_{23}v_\eta ) \\
(G_{31}v_{\sigma_2}-F_{13}v_\eta ) &
(G_{32}v_{\sigma_2}-F_{23}v_\eta ) &
G_{33}v_{\sigma_2}\,\,\,\,\,\,\,\,\,
\end{array}
\right ), \label{lepmass1} \eea
respectively.

The question we are going to pursue next is whether it is
possible to have some configuration where these interdependent
matrices assume a realistic form. In other words, if there exist
values of Yukawa couplings, $G_{ab}$ and $F_{ab}$, for which the
expected neutrino mixing as well as the charged lepton mass
spectrum can come about naturally.

We start by noticing that lepton mass matrices in
Eqs.~(\ref{neutmass1}) and (\ref{lepmass1}), can be diagonalized
as follows,
\be M_l^D = V_{eL}^\dag M_lV_{eR}\,,\,\,\,\,\,\, M_\nu^D =
V_{\nu}^\dag M_\nu V_{\nu}\,, \label{diagmass} \ee
where we have used the notation, $M^D\equiv Diag(m_1,m_2,m_3)$,
with $m_i$ being the physical lepton masses. The matrices $V$
transform the lepton fields in the interaction eigenstates into
mass eigenstates and, in principle, they are different for
left-handed and right-handed fields. These diagonalization
matrices $V_{eL}$ , $V_{eR}$ and $V_\nu$  combine themselves
 in the charged currents of the model
which, after the $3-3-1$ breaking to the $SU(3)_C \otimes
U(1)_{em}$, are given by:
\be {\cal L}^{CC}_{l}= -\frac{g}{\sqrt{2}}\left\{\overline{ e_L}
\gamma^\mu O^W\nu_LW^-_\mu + \overline{(e_R)^c}O^V\gamma^\mu
\nu_L V^+_\mu + \overline{(e_R)^c}O^U \gamma^\mu
e_LU^{++}_\mu\right\} + \mbox{H.c.}\,, \label{cc} \ee
where we are omitting family indices and $O^W=V^T_{eL} V_{\nu} $,
$O^V=V^T_{eR} V_{\nu}$ and $O^U=V^T_{eR} V_{eL}$ are the three
mixing matrices involved in the charged currents and all of them
can be chosen to be of the CKM type. Besides, we will be
considering the simplest case of a zero CP violating phase,
since this phase is irrelevant throughout our analysis. This
means that these matrices are real, and we can work with their
transpose instead of hermitian conjugate. The parameterization we
assume here for the mixing matrices is:
\bea O^{W,\,V,\,U}=  \left (
\begin{array}{ccccc}
c_{13} c_{12} &\,\,& s_{12} c_{13} &\,\,& s_{13} \\
-s_{12} c_{23}-s_{23} s_{13} c_{12} &\,\,& c_{23} c_{12}-s_{23}
s_{13} s_{12} &\,\,& s_{23} c_{13} \\s_{23} s_{12} -s_{13} c_{23}
c_{12} &\,\,&  -s_{23} c_{12}-s_{13} s_{12} c_{23}&\,\,& c_{23}
c_{13}
\end{array}
\right )\,, \label{CKM} \eea
where we have used the short form $c_{ij}\equiv
\cos{\theta_{ij}}$ and $s_{ij}\equiv \sin{\theta_{ij}}$.

From the three mixing matrices that form the lepton mixing in Eq.~(\ref{cc}),
we have experimental information only upon the angles of  $O^W$,  the mixing
matrix involved in the neutrino oscillation experiments. However the
information we have is not enough to determine uniquely the three angles in
Eq.~(\ref{CKM}). What we know is that the recent analysis of atmospheric
neutrino oscillation still favor $\nu_\mu - \nu_\tau$ oscillation with an
almost maximal mixing $0.92 < \sin^22\theta_{atm} \leq 1.0$ at 90 \% C.L.
\cite{atmospheric}. We also have that the oscillation among $\nu_e -\nu_\mu$ is
almost settled as the explanation for the  solar neutrino  problem. Here the 
recent results allow $0.25 \leq \sin^2 \theta_{sun}\leq 0.40$  and $0.6 \leq \cos^2 \theta_{sum} \leq 0.75$(90 \% C.L.)\cite{solar,peres}. Since the CHOOZ experiment failed to see the disappearance of
$\bar \nu_e$, we also have $0\leq \sin^2 2\theta_{chz} <0.1$ (90 \% C.L.)
\cite{chooz}.

For our proposal here we have to fix the angles $\theta_{atm}$
($\theta_{23}$ in Eq. (\ref{CKM})) and $\theta_{chz}$ ( $\theta_{13}$ in
Eq. (\ref{CKM})).  This
is straightforwardly done by taking the best fit for $\theta_{atm}
=45^{\circ}$, while $\theta_{chz}=0^{\circ}$, in agreement with
the above presented results. The angle involved in the solar
neutrino oscillation ($\theta_{12}$) is the one that allows
for a certain range of values. Fortunately, we can keep it as a
free parameter. For simplicity let us take $\sin \theta_{sun}=s$
and $\cos \theta_{sun}=c$. We then are left with the so called maximal mixing 
pattern for $O^W$, according to the parameterization in
Eq.~(\ref{CKM}):
\bea O^W= \left (
\begin{array}{ccc}
c & s & 0 \\
\frac{-s}{\sqrt{2}} & \frac{c}{\sqrt{2}} & \frac{1}{\sqrt{2}} \\
 \frac{s}{\sqrt{2}}& -\frac{c}{\sqrt{2}}  & \frac{1}{\sqrt{2}}
\end{array}
\right )\,. \label{neutmix} \eea
Henceforth we will focus our attention on this particular pattern
concerning the neutrino mixing.

\section{Possible lepton mixing }
\label{sec2}

According to Eq.~(\ref{lepmass1}), the charged lepton mass matrix
is necessarily non-diagonal, compelling us to analyze only two
cases. The first one is the case where neutrino mass matrix is
diagonal, and the mixing in the charged current is due only to
the charged lepton sector. The second possibility is to have both
neutrino and charged lepton mass matrices  non-diagonal, resulting in a 
 mixing matrix given by $O^W=V^T_{eL} V_{\nu}$.

\subsection{First Case}

Let us begin our analysis by the case where neutrinos are in a
diagonal mass basis. In this case, $O^W=V^T_{eL}$, $M_\nu =
Diag(G_{11}v_{\sigma_1},G_{22}v_{\sigma_1},G_{33}v_{\sigma_1})$
and $M_l$ takes the form
\bea M_l= \left (
\begin{array}{lcr}
G_{11}v_{\sigma_2} & F_{12}v_\eta & F_{13}v_\eta  \\
-F_{12}v_\eta  & G_{22}v_{\sigma_2} & F_{23}v_\eta  \\
-F_{13}v_\eta  & -F_{23}v_\eta  & G_{33}v_{\sigma_2}
\end{array}
\right )\,. \label{charglepmass} \eea
This texture is very peculiar since the non-diagonal elements are
anti-symmetric and, as far as we know, there is no approach
dealing with such a texture in literature. Now let us check if we
are able to obtain the mixing in Eq.~(\ref{neutmix}) taking into
account the texture in Eq.~(\ref{charglepmass}). It is suitable to
remark that when $V_\nu = I$ then $O^W=V^T_{eL}$ and
$O^V=V^T_{eR}$, meaning that $V^T_{eL}$ is given solely by the
maximal mixing matrix, Eq.~(\ref{neutmix}).

In this case $O^V=V^T_{eR}$. With this and using the pattern of a
CKM matrix in Eq.~(\ref{CKM}) for $O^V$, we have that
\bea V_{eR}= \left (
\begin{array}{ccccc}
c_{13} c_{12} &\,\,&  -s_{12} c_{23}-s_{23} s_{13} c_{12}&\,\,&
s_{23} s_{12} -s_{13}
c_{23} c_{12} \\
s_{12} c_{13} &\,\,& c_{23} c_{12}-s_{23} s_{13} s_{12} &\,\,&
-s_{23} c_{12}-s_{13}
s_{12} c_{23}  \\
s_{13} &\,\,& s_{23} c_{13} &\,\,& c_{23} c_{13}
\end{array}
\right )\,. \label{VeR} \eea

Starting with these assignments for the matrices $V_{eL}$ and
$V_{eR}$, the  charged lepton mass matrix,
Eq.~(\ref{charglepmass}), is connected to them through $M_l=V_{eL}
M^D_l V^T_{eR}$, where $M^D_l=Diag(m_e , m_\mu , m_\tau)$. Since
we know the charged lepton masses we can check if there exists a
range of values for the mixing angles that is compatible with this
equation, which we write as,
\bea V_{eL} M^D_l V^T_{eR}= \left (
\begin{array}{lcr}
M_{11} &M_{12} & M_{13} \\
M_{21} & M_{22} & M_{23}  \\
M_{31} & M_{32} & M_{33}
\end{array}
\right )\,, \label{clmtexture} \eea
where we have defined,
\bea &&M_{12}=
cc_{13}s_{12}m_e+\frac{s}{\sqrt{2}}[(s_{12}s_{13}s_{23}-c_{12}c_{23})m_\mu
-(c_{12}s_{23}+s_{12}s_{13}c_{23})m_\tau]\,,\nonumber \\
&&M_{21}=sc_{12}c_{13}m_e
-\frac{c}{\sqrt{2}}[(s_{12}c_{23}+c_{12}s_{13}s_{23})m_\mu
+(s_{12}s_{23}-c_{12}s_{13}c_{23})m_\tau]\,,\nonumber \\
&&M_{13}=cs_{13}m_e+\frac{s}{\sqrt{2}}[-c_{13}s_{23}m_\mu+
c_{13}c_{23}m_\tau]\,,
\nonumber \\
&&M_{31}=\frac{1}{\sqrt{2}}[-(s_{12}c_{23}+c_{12}s_{13}s_{23})m_\mu
+(s_{12}s_{23}-c_{12}s_{13}c_{23})m_\tau]\,,\nonumber \\
&&M_{23}=ss_{13}m_e+\frac{c}{\sqrt{2}}[c_{13}s_{23}m_\mu
-c_{13}c_{23}m_\tau]\,,
\nonumber \\
&&M_{32}=\frac{1}{\sqrt{2}}[(c_{12}c_{23}-s_{12}s_{13}s_{23})m_\mu
-(s_{12}s_{13}c_{23}+c_{12}s_{23})m_\tau ]\,,\nonumber \\
&&M_{11}=cc_{12}c_{13}m_e +\frac{s}{\sqrt{2}}[(s_{12}c_{23}
+c_{12}s_{13}s_{23})m_\mu +(s_{12}s_{23}
-c_{12}s_{13}c_{23})m_\tau]\,,\nonumber \\
&&M_{22}=ss_{12}c_{13}m_e+\frac{c}{\sqrt{2}}[(c_{12}c_{23}-
s_{12}s_{13}s_{23})m_\mu+(c_{12}s_{23}+s_{12}s_{13}c_{23})m_\tau]\,,\nonumber \\
&&M_{33}=
\frac{1}{\sqrt{2}}[c_{13}s_{23}m_\mu+c_{13}c_{23}m_\tau]\,.
\label{elements} \eea

According to Eqs.~(\ref{neutmass1}) and (\ref{lepmass1}), once we
impose the neutrino mass matrix is diagonal, the texture of the
charged lepton mass matrix automatically emerges and it is such
that the non-diagonal elements are anti-symmetric as displayed in
Eq.~(\ref{charglepmass}). This implies that the non-diagonal
elements in Eq.~(\ref{elements}) above have to be anti-symmetric,
$M_{ij}=-M_{ji}$ for $i\neq j$. These conditions can be
translated into the following set of non-linear equations for the
mixing angles:
\bea &&\sqrt{2}[cs_{12}c_{13} +sc_{12}c_{13}] m_e + [
s(s_{12}s_{13}s_{23}-c_{12}c_{23}) -
c(s_{12}c_{23}+c_{12}s_{13}s_{23})] m_\mu \nonumber \\
&& + [c(c_{12}s_{13}c_{23}-s_{12}s_{23})-s( c_{12}s_{23}
+ s_{12}s_{13}c_{23})]m_\tau=0\,,\nonumber \\
&&\sqrt{2}cs_{13}m_e -[sc_{13}s_{23} +
s_{12}c_{23}+c_{12}s_{13}s_{23}]m_\mu
+[sc_{13}c_{23}+s_{12}s_{23}-c_{12}s_{13}c_{23}]m_\tau=0\,,
\nonumber \\
&&\sqrt{2}ss_{13}m_e +[ cc_{13}s_{23}+
c_{12}c_{23}-s_{12}s_{13}s_{23}]m_\mu  -
[cc_{13}c_{23}+c_{12}s_{23}+s_{12}s_{13}c_{23}]m_\tau=0\,. \nonumber \\
\label{nonlineq} \eea

Motivated by the fact that the charged lepton masses differ by orders of magnitude
 the first naive attempt we could try in order to see if this
system admits a solution is to assume that the coefficients of
each mass term  vanishes. In the sequence we are going to
show that no solution can be found in this way and, in this case,
we have to allow for any combination of the mass coefficients to
cancel each other, which will need some numerical computation.

We start by making the coefficients of $m_e$, $m_\mu$ and
$m_\tau$ in Eq.~(\ref{nonlineq}), to vanish identically. This
leads automatically to the following constraints,
\be cs_{12}=-sc_{12}\,, \label{cs12} \ee
%
and
\be s_{13}=0\,. \label{s13} \ee
%

For simplicity, we are going to assume that
$c_{13}=+1$\footnote{It can be shown that $c_{13}=-1$ would lead
to the same conclusions, although to different charged lepton
mass elements in Eq.~(\ref{elements}).}. The constraint in
Eq.~(\ref{s13}) allows some simplification in Eq.~(\ref{nonlineq})
and we can extract further constraints from it,
namely,
\bea sc_{23}&=&-s_{12}s_{23}\,, \nonumber \\
ss_{23}&=&-s_{12}c_{23}\,, \nonumber \\
cc_{23}&=&-c_{12}s_{23}\,, \nonumber \\
cs_{23}&=&-c_{12}c_{23}\,. \label{false} \eea
Taking into account the constraint Eq.~(\ref{cs12}) together with
these equations, it is straightforward to see that we only have a
solution for this system if $s=0$ or $c=0$, which is not the case
since the small angle MSW solution is already ruled out. In this way we
have shown that our naive analytical approach does not allow us
to obtain a conclusion about the existence of a solution.

We have then to resort to numerical calculations. The numerical
a\-na\-ly\-sis
was done using the package for solving systems of
non-linear equations described in Ref.~\cite{NLS}. We fixed the
lepton masses as, $m_e=0.51$~MeV, $m_\mu =105.66$~MeV and $m_\tau
=1777.0$~MeV, and assumed values for $s$ and $c$ in the range:
$0.439941< s\simeq c <0.686221$~\cite{BahcallConcha}. No solution
for the system Eq.~(\ref{nonlineq}) was found in this range. This
analysis is sufficient to conclude that lepton mixing in minimal
3-3-1 model does not allow for the pattern of charged lepton
masses given by Eq.~(\ref{charglepmass}). Hence, we should look
for another scenario where neutrinos and charged lepton mass
matrices are both in a non-diagonal basis. That is what we will do next.

\subsection{Second Case}

Another possible scenario is one  where both neutrino and
charged lepton mass matrices are non-diagonal. In this situation,
the mixing in Eq.~(\ref{neutmix}) is generated partially by the
diagonalization of neutrino mass matrix and partially by the
diagonalization of charged lepton mass matrix. The only way of
separating Eq.~(\ref{neutmix}) in these two matrices is to have
maximal mixing between $\nu_\mu$ and $\nu_\tau$ coming from the
charged lepton sector and the mixing in the $\nu_e$ to $\nu_\mu$
oscillation coming from the neutrino sector. In this way we have
\bea O^W=V^T_{eL} V_{\nu}= \left (
\begin{array}{ccc}
c & s & 0 \\
\frac{-s}{\sqrt{2}} & \frac{c}{\sqrt{2}} & \frac{1}{\sqrt{2}} \\
 \frac{s}{\sqrt{2}}& -\frac{c}{\sqrt{2}}  & \frac{1}{\sqrt{2}}
\end{array}
\right )=\left (
\begin{array}{ccc}
1 & 0 & 0 \\
0 & \frac{1}{\sqrt{2}} & \frac{1}{\sqrt{2}} \\
 0& -\frac{1}{\sqrt{2}}  & \frac{1}{\sqrt{2}}
\end{array}
\right )\times \left (
\begin{array}{ccc}
 c &  s  & 0 \\
 -s &  c  & 0 \\
 0 &  0  & 1
\end{array}
\right )\,, \label{division} \eea
where,
\bea V^T_{eL} =\left (
\begin{array}{ccc}
1 & 0 & 0 \\
0 & \frac{1}{\sqrt{2}} & \frac{1}{\sqrt{2}} \\
 0& -\frac{1}{\sqrt{2}}  & \frac{1}{\sqrt{2}}
\end{array}
\right ), \label{clepton} \eea
and,
\bea
 V_{\nu}= \left (
\begin{array}{ccc}
c & s & 0 \\
-s & c & 0 \\
 0& 0  & 1
\end{array}
\right )\,. \label{neutpart} \eea
From the Lagrangian term containing the singly charged bilepton,
Eq.~(\ref{cc}), we have that $O^V=V^T_{eR} V_{\nu}$, with the
diagonalization matrices given above. We then parameterize $O^V$
according to the usual CKM pattern,
\bea O^V= \left (
\begin{array}{ccccc}
c_{12} c_{13} &\,\,& s_{12} c_{13} &\,\,& s_{13}  \\
-s_{12} c_{23}- c_{12} s_{13} s_{23} &\,\,& c_{12} c_{23}-s_{12}
s_{13} s_{23} &\,\,&
c_{13} s_{23} \\
s_{12} s_{23} -c_{12}s_{13}c_{23}&\,\,& -c_{12} s_{23}-s_{12}
s_{13} c_{23} &\,\,& c_{13} c_{23}
\end{array}
\right ) \, . \label{VeRa} \eea
Dissociating it as a product of three different rotation
matrices,
\bea O^V= \left (
\begin{array}{ccc}
1 & 0 & 0  \\
0 &  c_{23} & s_{23}  \\
0 & -s_{23} & c_{23}
\end{array}
\right )\times \left (
\begin{array}{ccc}
c_{13} & 0 & s_{13}  \\
0 &  1 & 0  \\
-s_{13} & 0 & c_{13}
\end{array}
\right ) \times \left (
\begin{array}{ccc}
c_{12} & s_{12} & 0  \\
-s_{12} &  c_{12} & 0 \\
0 & 0 & 1
\end{array}
\right ), \label{VeR1} \eea
we  recognize  the last matrix at the right hand side of
Eq.~(\ref{VeR1}) as  the neutrino mixing  matrix given above in
Eq.~(\ref{neutpart}) which implies $c_{12}=c$  and $s_{12}=s$.

After this,  the appropriate form of $V_{eR}$ is then
automatically obtained as follows:
\bea V^T_{eR}= \left (
\begin{array}{ccc}
1 & 0 & 0  \\
0 &  c_{23} & s_{23}  \\
0 & -s_{23} & c_{23}
\end{array}
\right )\times \left (
\begin{array}{ccc}
c_{13} & 0 & s_{13}  \\
0 &  1 & 0  \\
-s_{13} & 0 & c_{13}
\end{array}
\right ) =\left (
\begin{array}{ccc}
c_{13} & 0 & s_{13}  \\
-s_{23}s_{13} &  c_{23} & s_{23}c_{13}  \\
-c_{23}s_{13} & -s_{23} & c_{23}c_{13}
\end{array}
\right )\,, \label{VeR2} \eea
With $V_{eR}$ given above, Eq.~(\ref{VeR2}), and $V_\nu$ given by
Eq.~(\ref{neutpart}), we are able to obtain the texture of the
neutrino and charged lepton mass matrices. For the neutrinos we
have:
\bea V_\nu M_\nu^D V^T_\nu = \left (
\begin{array}{ccccc}
m_1 c^2 + m_2 s^2 &\,\,& (m_2-m_1) cs &\,\,& 0  \\
(m_2-m_1) cs &\,\,&  m_1 s^2 + m_2 c^2 &\,\,& 0  \\
0 &\,\,& 0 &\,\,& m_3
\end{array}
\right )\,. \label{neumass2} \eea
When this is confronted with $M_\nu$ in Eq.~(\ref{lepmass1}) we
conclude that $G_{13}=G_{23}=0$. As a consequence, we are left
with a peculiar texture for the charged lepton mass matrix,
\bea M_l= \left (
\begin{array}{ccc}
G_{11}v_{\sigma_2} & (G_{12}v_{\sigma_2}+F_{12}v_\eta )&
F_{13}v_\eta  \\
(G_{12}v_{\sigma_2}-F_{12}v_\eta ) & G_{22}v_{\sigma_2} &
F_{23}v_\eta  \\
-F_{13}v_\eta  & -F_{23}v_\eta  & G_{33}v_{\sigma_2}
\end{array}
\right )\,. \label{lepmass2} \eea
We must check now if these assignments are plausible by verifying
if the diagonalization matrices so obtained drive us to such a
texture. We then compute $V_{eL}M^D_l V^T_{eR}$ where $V_{eL}$ is
given by Eq.~(\ref{clepton}), and $V_{eR}$ by Eq.~(\ref{VeR2}),
\bea&& M_l=V_{eL}M^D_l V^T_{eR}= \left (
\begin{array}{ccc}
1 & 0 & 0  \\
0 &  \frac{1}{\sqrt{2}} & -\frac{1}{\sqrt{2}}  \\
0 & \frac{1}{\sqrt{2}} & \frac{1}{\sqrt{2}}
\end{array}
\right )\times \left (
\begin{array}{ccc}
m_e & 0 & 0  \\
0 &  m_\mu & 0  \\
0 & 0 & m_\tau
\end{array}
\right ) \times \left (
\begin{array}{ccc}
c_{13} & 0 & s_{13}  \\
-s_{23}s_{13} &  c_{23} & s_{23}c_{13}  \\
-c_{23}s_{13} & -s_{23} & c_{23}c_{13}
\end{array}
\right )= \nonumber \\ && \nonumber \\ && \left (
\begin{array}{ccccc}
c_{13}m_e &\,\,& 0 &\,\,& s_{13}m_e  \\
\frac{1}{\sqrt{2}}(-s_{23}s_{13}m_\mu + c_{23}s_{13}m_\tau)&\,\,&
\frac{1}{\sqrt{2}}( c_{23}m_\mu +s_{23}m_\tau)&\,\,&
\frac{1}{\sqrt{2}}(s_{23}c_{13}m_\mu
-c_{23}c_{13}m_\tau)  \\
\frac{1}{\sqrt{2}}(-s_{23}s_{13}m_\mu-c_{23}s_{13}m_\tau) &\,\,&
\frac{1}{\sqrt{2}}(c_{23}m_\mu-s_{23}m_\tau) &\,\,&
\frac{1}{\sqrt{2}}(s_{23}c_{13}m_\mu+c_{23}c_{13}m_\tau)
\end{array}
\right )\,. \label{massleptext} \eea

Let us look at the consequences of this result. We first observe
that by identifying Eq.~(\ref{lepmass2}) with
Eq.~(\ref{massleptext}) we obtain the following equations,
\bea && -s_{23}s_{13}m_\mu + c_{23}s_{13}m_\tau = 0\,,\nonumber \\
&& s_{13}m_e = \frac{1}{\sqrt{2}}(s_{13}s_{23}m_\mu +
s_{13}c_{23}m_\tau )\,,\nonumber \\
&& c_{23}m_\mu-s_{23}m_\tau = -s_{23}c_{13}m_\mu +
c_{23}c_{13}m_\tau\,. \label{eqnond} \eea
We can analyze two possibilities, namely, $s_{13}=0$ and
$s_{13}\neq 0$. For the case $s_{13}=0$ (assuming $c_{13}=+1$),
these equations have a solution for $c_{23}=-s_{23}$.

However, the fact that elements $G_{ab}$ are present in both mass
matrices, of neutrinos and charged leptons, leads to tight
constraints on their values. For instance, by comparing the matrix
in Eq.~(\ref{lepmass1}) with Eq.~(\ref{neumass2}), we can extract
\be G_{11}v_{\sigma_1}=m_1 c^2 + m_2 s^2\,,\label{g11} \ee
and
\be G_{33}v_{\sigma_1}=m_3\,.\label{g33} \ee
According to Eqs.~(\ref{lepmass2}) and (\ref{massleptext}), after
substituting the obtained value for the mixing angles, $s_{13}=0
\Rightarrow c_{13}=1$ and $c_{23}=-s_{23}=1/\sqrt{2}$, $G_{11}$
and $G_{33}$ are also constrained to the charged lepton masses,
\be G_{11}v_{\sigma_2}=m_e\,,\label{g11m} \ee
and
\be G_{33}v_{\sigma_2}=\frac{1}{2}(m_\tau - m_\mu)\simeq
\frac{1}{2}m_\tau\,. \label{g33m} \ee
But this is not possible because $m_1$, $m_2$ and $m_3$ are of the
order of eV, while $m_e$ is of the order of MeV and $m_\tau$ is
of the order of GeV. When we divide Eq.~(\ref{g11}) and
Eq.~(\ref{g33}) by Eq.~(\ref{g11m}) and Eq.~(\ref{g33m}),
respectively, we obtain that the ratio
$v_{\sigma_1}/v_{\sigma_2}$ is about $10^{-6}$ in one case and
$10^{-9}$ in the other, which is an absurd. This shows that such a
solution is not possible.

Suppose now that $s_{13}$ is not null. In this situation the
second equation in Eq.~(\ref{eqnond}) becomes,
\be m_e = \frac{1}{\sqrt{2}}(s_{23}m_\mu + c_{23}m_\tau )\,.
\label{eqme} \ee
Since the magnitude of the charged lepton masses are very
different, there is no value for the mixing angle in
Eq.~(\ref{eqme}) that makes the right hand side of this equation
to be equal to the electron mass. For instance, if
$s_{23}\,(c_{23})$ is small, $c_{23}\,(s_{23})$ is close to one
and we would obtain $m_e\sim m_\tau\,(m_\mu)$, an obviously false
result. This analysis is general enough and we do not even need to
appeal to numerical computation to conclude that the minimal 3-3-1
model requires some extension in order to accommodate the lepton
mixture.

\section{Concluding Remarks}
\label{sec3}

We have investigated the possibility of conciliating the mixing
in lepton sector, suggested by the neutrino oscillation
hypothesis, with the minimal version of 3-3-1 model. Due to the
very particular form of the anti-symmetric Yukawa coupling with
the triplet, we have shown that none of the alternatives for the
mass matrix textures of charged leptons and neutrinos could
generate the expected pattern for neutrino mixing. Hence, if by
one side the minimal 3-3-1 model naturally explains the smallness
of neutrinos masses~\cite{joshi}, by the other side it is not
able to consistently accommodate the neutrino mixing. This does
not mean that the model loses its attractiveness, it still
addresses additional questions not embraced by the Standard Model.
However, the newest observations from neutrino physics put it
clear, through the results here obtained, that further extension
is needed in order to keep the privileged position that the model
deserves\cite{yasue}.

{\it Acknowledgements.}  
We thank Orlando Peres for valuable informations on the recent experimental
limit on $\sin^2 \theta_{sum}$  and $\cos^2 \theta_{sum}$. A.G  and P.R.S  thank
 FAPESP while C.P thank CNPq for partial support.


\begin{thebibliography}{99}

\bibitem{viceframp} F.~Pisano and V.~Pleitez, Phys. Rev. D{\bf 46}, 410
(1992); P.~H.~Frampton, Phys. Rev. Lett. {\bf 69}, 2889 (1992).


\bibitem{footpisano} R.~Foot, O.~F.~Hern\'andez, F.~Pisano and V.~Pleitez,
Phys. Rev. D{\bf 47}, 4158 (1993).


\bibitem{SNO} SNO collaboration, Q.R. Ahmad {\it et al.},
Phys. Rev. Lett. {\bf 89}, 011301 (2002); Phys. Rev. Lett. {\bf 89}, 011302 (2002).

\bibitem{K2K} Super-Kamiokande Collaboration (Y.~Fukuda {\it et
al.}), Phys.~Lett. B{\bf 436}, 33 (1998); Phys.~Rev.~Lett. {\bf
81}, 1158 (1998); Erratum {\bf 81}, 4279 (1998); Phys.~Rev.~Lett.
{\bf 81}, 1562 (1998); Phys.~Rev.~Lett. {\bf 82}, 1810 (1999);
Super-Kamiokande collaboration, Y. Suzuki, Nucl.~Phys. B{\bf 91}
(Proc.~Suppl.), 29 (2001); Super-Kamiokande collaboration, S.
Fukuda et al., Phys.~Rev.~Lett. {\bf 86}, 5651 (2001).

\bibitem{KamLAND} KamLAND Collaboration, K. Eguchi {\it et al.}, Phys. 
Rev. Lett. {\bf 90}, 021802 (2003).

\bibitem{BahcallConcha} J.~N.~Bahcall, M.~C.~Gonzalez-Garcia and
C.~Pena-Garay, JHEP {\bf 0204}, 007 (2002); JHEP {\bf 0302}, 009 (2003).

\bibitem{joshi} J. C. Montero, C. A. de S. Pires and  V. Pleitez, 
 Phys. Lett. B{\bf 502}, 167, (2001);
  M.~B.~Tully and G.~C.~Joshi, Phys.~Rev. D{\bf 64},
011301(R) (2001).

\bibitem{atmospheric} G.~L. Fogli, E.~Lisi, A.~Marrone, and D.~Montanino,  hep-ph/0303064. 

\bibitem{solar} G.~L.~Fogli, E.~Lisi, A.~Marrone, D.~Montanino, A. Palazzo, and  A.~M.~Rotunno, hep-ph/0212127.

\bibitem{peres} 
H. Nunokawa, W. J. C. Teves, R. Zukanovich Funchal,
Phys. Lett. B {\bf 562}, 28 (2003); 
P. C. de Holanda, A. Y. Smirnov, JCAP {\bf 0302}, 001 (2003).
 


\bibitem{chooz}
CHOOZ Collaboration, M. Apollonio {\it et al.}, Phys. Lett. B {\bf 420}, 
397 (1998).


\bibitem{NLS} R.~B.~Kearfott and M.~Novoa, ACM Trans. Math.
Software {\bf 16}, 152 (1990).


\bibitem{yasue} For various approaches in this directions see:
Y. Okamoto and M. Yasue,  Phys. Lett. B {\bf 466} 267 (1999); 
T. Kitabayashi and  M. Yasue,  Nucl. Phys. B {\bf609} 61 (2001) and references therein;
J. C. Montero, C. A. de S. Pires and  V. Pleitez,
 Phys. Rev. D{\bf 65} 095001 (2002); J. C. Montero, C. A. de S. Pires, and  V. Pleitez,  Phys. Rev. D{\bf 66} 113003 (2002);
 J. C. Montero, C. A. de S.Pires, and  V. Pleitez,  Phys. Rev. D{\bf 65} 093017 (2002). 
 



\end{thebibliography}
\end{document}